\documentclass[usenatbib,authoryear,preprint,12pt]{elsarticle}

\usepackage{epsfig}
\usepackage{amssymb}
\usepackage{graphicx}
\usepackage{subfigure}
\usepackage{rotating}
\newcommand{\de}{{\rm d}}
\newcommand{\bea}{\begin{eqnarray}}
\newcommand{\eea}{\end{eqnarray}}
\newcommand{\f}{\frac}
     
\begin{document}
\begin{frontmatter}

\title{Models of high redshift luminosity functions 
and galactic outflows: The dependence on halo mass function
}
\author{Saumyadip Samui\corref{cor1}}
\ead{samui@iucaa.ernet.in}
\cortext[cor1]{IUCAA, Post Bag 4, Ganeshkhind, Pune 411 007, India,
Phone:+91-20-25604219, Fax:+91-20-25604699}
\author{Kandaswamy Subramanian}
\ead{kandu@iucaa.ernet.in}
\author{Raghunathan Srianand}
\ead{anand@iucaa.ernet.in}

\address{IUCAA, Post Bag 4, Ganeshkhind, Pune 411 007, India.}

\begin{abstract}
The form of the halo mass function is a basic ingredient in
any semi-analytical galaxy formation model. We study the
existing forms of the mass functions in the
literature and compare their predictions for semi-analytical
galaxy formation models. Two methods are used in the literature to
compute the net formation rate of halos,
one by simply taking the derivative of the halo mass function 
and the other using the prescription due to Sasaki (1994). 
For the historically used Press-Schechter (PS) mass
function, we compare various model predictions,
using these two methods. However, as the Sasaki formalism
cannot be easily generalized for other mass functions,
we use the derivative
while comparing model predictions of different
mass functions. 
We show that the reionization history and UV luminosity function of Lyman break
galaxies (LBGs) predicted by the PS mass
function differs from those using any other existing mass function, like
Sheth-Tormen (ST) mass function. 
{ In particular the reionization efficiency of molecular
cooled halos has to be substantially reduced when one uses the
ST and other mass functions obtained from the simulation
instead of the PS mass function.
Using $\chi^2$-minimization, we find that the observed 
UV luminosity functions of LBGs at $3.0\le z\le 7.4$ are
better reproduced by models using the ST mass function
compared to models that use the PS mass function.}
On the other hand, 
the volume filling factor of the metals expelled from the galaxies through
supernovae driven outflows differs very
little between models with different mass functions. It depends 
on the way we treat merging outflows.
We also show that the
porosity weighted average quantities related to the outflow are not very
sensitive to the differences in the halo mass function.
\end{abstract}
 
 
\end{frontmatter}

\section{Introduction}

The formation of galaxies and their influence on the physical state
of the inter-galactic medium (IGM) is often studied using
semi-analytical models of structure formation
(for example 
\citealt{WF91,kauffmann1993,cole94,SP99,chiu,madau_ferrara_rees,somerville,scannapieco}).
Such studies 
will allow one to extensively explore the unknown parameter space related
to physical conditions in a high redshift
galaxy and its star formation, with limited computational
resources. Semi-analytical models together
with available observations can be used to understand
the evolution of our universe while putting tight
constraints on various physical processes associated with
the high redshift universe.
In Samui et al. (2007, 2008) (hereafter Paper~I and Paper~II)
we used semi-analytical models
to predict the ionization history
of the universe, the high redshift ($z\ge 3 $) UV luminosity 
functions of LBGs and also the global effect of galactic outflows.

In any semi-analytical galaxy formation model, one needs to
assume some form of the halo mass function giving the
number density of dark matter halos
as a function of mass and redshift. The
first halo mass function in an analytical form was given
by Press \& Schechter (1974). The Press-Schechter (PS)
mass function has been widely employed in semi-analytical
models of galaxy formation and in understanding the high
redshift universe.  It is important to note that 
one needs not just the dark halo mass function
at any redshift, but also their formation rate and survival probability.
For the PS mass function, Sasaki (1994) proposed
a method to find the formation rate of halos and their
survival probability at later epochs. 
Note that one can also find this rate through N-body simulations,
or by constructing a large number of merger trees.
However, it turns out that we need to resolve
halo masses 
over a wide dynamic range from $10^7M_\odot$ to $10^{12}M_\odot$,
to model both the effect of outflows and the galactic luminosity function.
One also needs to have a large enough number of halos in the simulation
or a large number of realizations of the merger trees, 
to obtain good statistics of the range of galaxy properties.
Therefore, to have a preliminary look at this challenging
problem, in Paper~I and ~II,
we focused on analytical methods. In particular we used 
the PS mass function
and the Sasaki formalism, to predict the ionization history
of the universe, the high redshift ($z\ge 3 $) UV luminosity 
functions and also the global effect of galactic outflows.

In Paper~I, we concluded that the observed redshift evolution of
the luminosity function of Lyman break galaxies (LBGs) at $z\ge 3$
requires evolution in physical properties that govern the
star formation activity on top of the evolution in the number density of
dark matter halos coming from the structure formation model.
We found that the required amount of redshift evolution in the
star formation activities depends on the redshift
evolution of the number density of dark matter halos and hence
on the assumed form of the halo mass function.

In Paper~II, we showed that galactic scale outflows
originating predominantly 
from small mass halos with mass
$10^7 - 10^9~M_\odot$ can pollute 60\%
of the IGM with metals at $z\sim 3$ with
a metallicity floor of $10^{-3}Z_\odot$. Inclusion of star formation in even
smaller mass halos which cool due to the presence of H$_2$ or HD
molecules, helps more in spreading the metals. This is because they
are more abundant and hence their mean separation is smaller. In these
models, even at $z=8$, more than 60\% of IGM is being polluted with
metals.
These conclusions may crucially depend on the choice of the
adopted form of the mass function.

It is known that the PS mass function does not provide a good fit
to the number density of dark matter halos determined from 
recent high resolution simulations of galaxy formation.
From such simulations a number of different fitting formulae
for the halo mass function have been suggested
(see for example Sheth \& Tormen 1999; Jenkins et al. 2001;
Reed et al. 2003, 2007 and Warren et al. 2006).
Note that there is no direct way to scale the results
obtained by assuming the PS mass function to results
which would arise from these other mass functions.
Therefore it is important to revisit various issues
discussed in Paper~I and II considering the mass functions
determined from the simulations. This forms the basic
motivation of this paper. 
We will mainly concentrate, as in Paper~I and II,
on high redshift (i.e. $z \ge 3$) UV luminosity function of
Lyman break galaxies (LBGs) and the effect of galactic outflows on the IGM
adopting a self consistent reionization history.

Note that the Sasaki formalism, used in Paper I and II,
is not easily generalisable to other mass functions
(Ripamonti, 2007 and also see discussion in section~2).
%,
An alternative is to simply take the derivative
of the mass function under the assumption that
it is sufficiently close to the net formation rate of objects in the mass
range that is of interest to the semi-analytic models.
This approach is widely used in several semi-analytical models in the
literature (for example Haehnelt \& Rees 1993, Scannapieco 2005).
The derivative includes not only the 
formation rate of halos, but also their destruction
rate at the same redshift. 
Therefore it is not guaranteed to be positive definite.
This is the disadvantage of using the derivative
of the mass function.
Nevertheless, in the range of halo masses that we will be
interested in at different redshifts, we will see that the
derivative of the mass function is indeed positive definite.
Therefore, for other mass functions consider in this work, we simply
use its derivative to model the net formation rate of dark matter halos.
We will also compare results obtained using 
the derivative as an alternative to the Sasaki formalism for the PS mass function.

The paper is organized as follows. In section~2 we review
different proposals for the halo mass function that we wish to study
and also compare the net formation rate of collapsed halos predicted
by these mass functions.
The star formation and reionization models are described in section~3.
In section~4 we show the effect of the
halo mass function on the predicted UV luminosity
functions of high redshift LBGs. 
{In particular, we use the $\chi^2$-minimization technique
to discriminate between models using different mass functions.}
The feedback of galactic outflows on
the IGM is discussed in section~5. Finally section~6 gives
our conclusions.
In this work we use the cosmological parameters consistent with the
recent WMAP 5th year data release i.e. $\Omega=1$, $\Omega_m = 0.26$,
$\Omega_\Lambda = 0.74$, $\Omega_b=0.044$, $h = 0.72$, $\sigma_8=0.80$
and $n_s =0.96$ (Dunkley et al. 2008).
Also we use a Salpeter stellar initial mass function (IMF)
in the mass range $1-100~M_\odot$ unless otherwise mentioned.

\section{Comparing halo mass functions}
The halo mass function is defined to be the number density
of collapsed dark matter halos in the mass range $M$ and $M+\de M$
at a given redshift $z$.
The differential halo mass function is defined as 
\begin{equation}
\f {\de N}{\de M} = \f{\rho_0}{M} \f{\de \ln \sigma^{-1}}{\de M}
  f(\sigma)
\label{eqn_mf}
\end{equation}
where $\rho_0$ is the average density of the universe at that redshift.
The rms density fluctuations, $\sigma$, of the smooth density field with
a top-hat window function is defined as
\begin{equation}
\sigma^2 = D^2(z) \sigma_0^2 = D^2(z) \f{1}{2\pi ^2}\int\limits_0^{\infty}
k^2 P(k)W^2(k,M) \de k
\end{equation}
where, $P(k)$ is the linear power spectrum of the density fluctuations
at $z = 0$, $W(k,M)$ is the Fourier transform of the real-space top-hat
filter, and $D(z)$ is the growth factor of linear perturbations normalized
to unity at $z = 0$ (Peebles 1993, Padmanabhan 2002). The idea of representing the mass
function in the form of Eq.~\ref{eqn_mf} is that different analytical
forms of the halo mass function can be represented with different forms of the
function $f(\sigma)$ in that equation. For example, $f(\sigma)$ for the PS
mass function is given by
\begin{equation}
f_{PS}(\sigma) = \sqrt{\f{2}{\pi}}\f{\delta_c}{\sigma} \exp{\left(
-\f{\delta_c^2}{2\sigma^2}\right)}
\label{eqn_PS}
\end{equation}
where $\delta_c$ is the critical over density for collapse,
usually taken to be equal to $1.686$.
The simple form of PS mass function has been widely used in the
semi-analytical models of galaxy formation (Chiu \& Ostriker 2000, Barkana \& Loeb 2001,
Nagamine et al 2006). 

The deviation of the PS mass function from numerical
simulations was pointed out by Sheth \& Tormen (1999). The
discrepancy is larger for high mass rare objects, the PS mass function always
predicting a smaller number of rare objects compared to numerical
simulations. Sheth \& Tormen (1999)
proposed a modification of the PS mass function which provides
a better fit to the numerical simulation data.
The Sheth-Tormen (ST) mass function takes the following form:
\begin{equation}
f_{ST}(\sigma)= A \sqrt{\f{2 a}{\pi}}\left[1 + \left(\f{\sigma^2}{a\delta_c^2}\right)^p
\right] \f{\delta_c}{\sigma}\exp{\left(-\f{a\delta_c^2}{2\sigma^2}\right)}.
\label{eqn_ST}
\end{equation}
Choosing $A = 0.3222$, $a = 0.707$ and $p = 0.3$ in
Eq.~\ref{eqn_ST} provides a better fit to mass functions obtained
from numerical simulations over a wide range of masses and redshifts
compared to the PS formula. Note that, if we take $A = 0.5$, $a = 1.0$ and $p = 0.0$
in Eq.~\ref{eqn_ST} then the ST mass function reduces to the PS mass function.

Further, Jenkins et al. (2001) found deviation from the ST mass function
in their simulations of the $\tau$CDM and $\Lambda$CDM
cosmologies. They proposed another $f(\sigma)$ 
which fits better their own simulations
over more than four orders of magnitude in mass,
$\sim3\times10^{11}$ to $\sim5 \times 10^{15}~h^{-1}M_\odot$.
This mass function, referred to here as Jenkins \& White (JW)
mass function is given by
\begin{equation}
f_{JW} = 0.315 \exp{\left[-\vert\ln\sigma^{-1} + 0.61\vert ^{3.8}\right]}.
\label{eqn_JW}
\end{equation}

Another study of $f(\sigma)$ was made by Reed et al.(2003).
They used high-resolution $\Lambda$CDM numerical
simulations to calculate $f(\sigma)$ of dark matter halos
down to the scale of dwarf galaxies, back to a redshift of $15$.
They showed that the ST mass function provides a good fit to 
their data except for redshift $10$ or higher where it over predicts
halo numbers by more than $50\%$.
In a later work, Reed et al. (2007) developed a new method for compensating
the effects of finite simulation volume. This allowed them to find an
approximation to the true ‘global’ mass function.
They proposed another fitting formula given by
\begin{eqnarray}
f_{R}(\sigma) &=& A \sqrt{\f{2a}{\pi}}\left[ 1 + \left( \f{\sigma^2}{a\delta_c^2}
\right)^p + 0.6G_1 + 0.4G_2\right]\f{\delta_c}{\sigma}\nonumber\\
& & \exp{\left[-\f{c a \delta_c^2}{2\sigma^2} - \f{0.03}{(n_{eff}+3)^2}
\left(\f{\delta_c}{\sigma}\right)^{0.6}\right]}
\label{eqn_Reed07}
\end{eqnarray}
where
\begin{eqnarray}
G_1 & = & \exp{\left[- \f{[\ln \sigma^{-1} - 0.4]^2}{2(0.6)^2}\right]}
\label{eqn_G1} 
\\
G_2 & = & \exp{\left[- \f{[\ln \sigma^{-1} - 0.75]^2}{2(0.2)^2}\right]}
\label{eqn_G2}
\end{eqnarray}
and
\begin{equation}
\f{n_{eff} + 3 }{6} = \f{\de \log \sigma^{-1}}{\de \log M}
\label{eqn_n}
\end{equation}
The values of constants which require to fit the numerical
simulations are $c = 1.08$, $ca = 0.764$, and $A = 0.3222$.

Meanwhile Warren et al. (2006) performed simulations
where they corrected a systematic error in halo-mass determination
by the friends-of-friends halo finder. They also measured the shape
and quantified the uncertainty in the predicted theoretical mass function
of dark matter halos in $\Lambda$CDM cosmology and showed that
the canonical ST and JW forms of the mass function are
inconsistent with the $\Lambda$CDM mass function at a $\sim 10\%$
level at intermediate masses, and $>30\%$ at the highest masses.
They provided an analytical fit to $f(\sigma)$ 
given by 
\begin{equation}
f_{W}(\sigma) = 0.7234 \left(\sigma^{-1.625} + 0.2538\right)
\exp{\left[ - \f{1.1982}{\sigma^2}\right]}.
\label{eqn_W}
\end{equation}

Thus we have a set of halo mass functions which gives a better fit
to the numerical simulations of galaxy formation compared
to the classical PS formula. Here, we
consider these set of $f(\sigma)$ to examine how sensitive
the predictions of semi-analytical models are to the 
assumed form of the halo mass function.

We begin by comparing the net formation rate of dark matter halos for
different halo mass functions described above. This is the most fundamental ingredient in any
semi-analytical model of galaxy formation. In Paper~I~\&~II
we used Sasaki formalism of the PS mass function to get the formation
rate. Following the same steps for the ST mass function we obtain
\begin{eqnarray}
{\cal N}_s(M,z,z_c) &=  &\f{\rho_0}{M}\left(-\f{1}{\sigma}\f{\de \sigma}{\de M}
\right) f_{ST}(\sigma)\left(\f{1}{D(z_c)}\f{\de D(z_c)}{\de z_c}\right) \nonumber \\
& & \left[ \f{\de \ln f_{ST}(\sigma)}{\de \ln \sigma} + ( 1 - 2p)\right]
\left[\f{D(z_c)}{D(z)}\right]^{1-2p} 
\label{eqn_ST_sasaki}
\end{eqnarray}
where ${\cal N}_s(M,z,z_c)$ is the formation rate of the dark matter
halos collapsed at $z_c$ and survive till some observed redshift, $z\le z_c$.
The survival probability is given by $[{D(z_c)}/{D(z)}]^{(1-2p)}$.
Note that putting $A = 0.5$, $a = 1.0$ and $p = 0$ in above equation
gives back the Sasaki formalism of the PS mass function (see eq.~1 of Paper~I).
It is clear from the above equation that due to the presence of $-2p$ term,
it is not guaranteed that the formation rate will always be positive.
As formation rate being negative is unphysical the generalization of
Sasaki formalism for the ST mass function is incorrect.
Therefore,  as mentioned in section 1,
we rather model the net formation rate of collapsed dark matter
halos, by taking the redshift derivative of the halo mass
function, ${\cal N}(M, z_c) = \de ^2N/ \de M \de z_c$.
\begin{figure*}
\centerline{%
\includegraphics[bb=55 3 600 580,width=1.\textwidth]{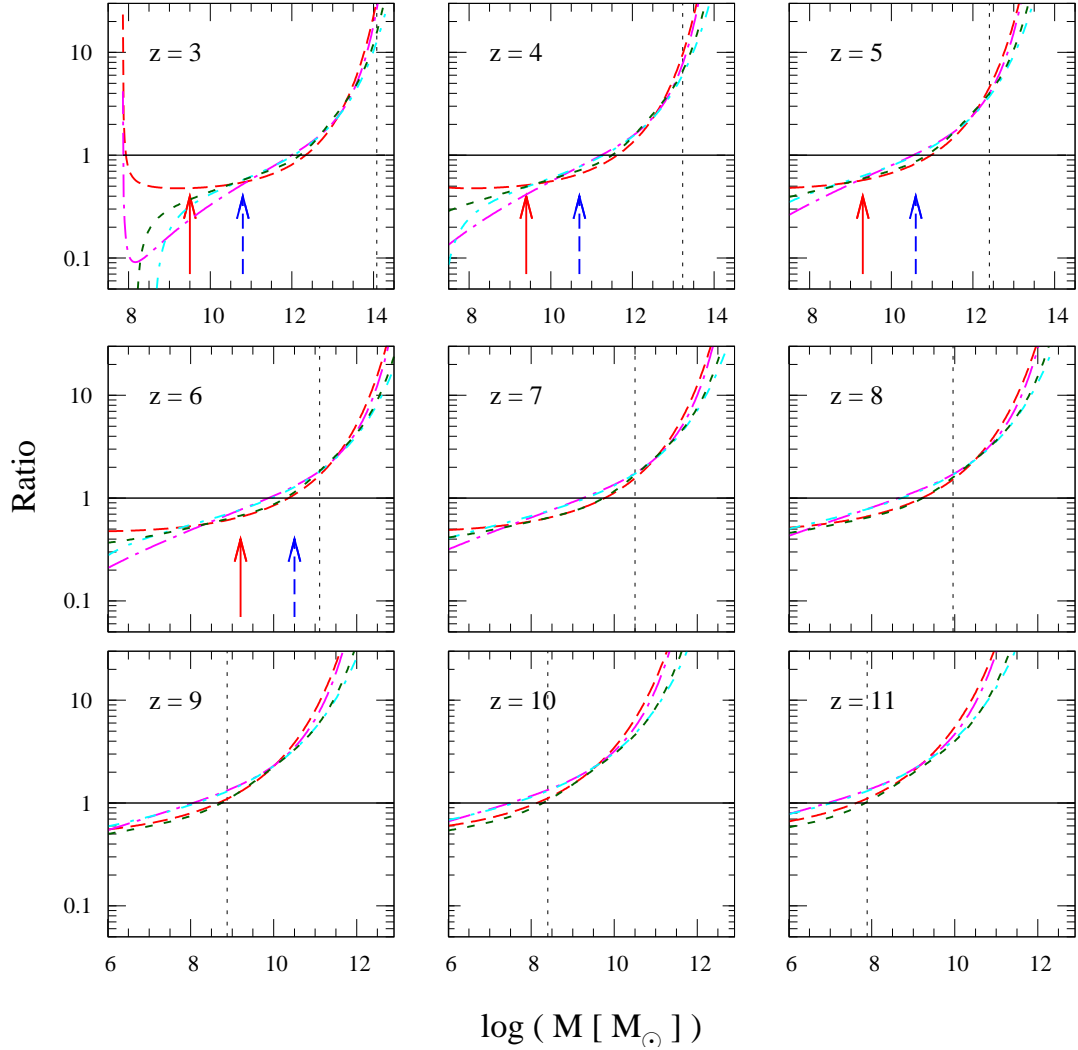}
}
\caption[]{The net formation rate of dark matter halos, ${\cal N}$, for different
halo mass functions at different redshifts. 
We compare the redshift derivative of different mass functions with the derivative
of the PS mass function (shown by the solid line).
We show the derivative of ST (long dashed, red), JW(dot long dashed, magenta),
Warren (dot short dashed,
cyan) and Reed (short dashed, dark green) mass functions. The vertical dashed line
in each panel shows the mass of typical collapsing objects from $3\sigma$ fluctuations.
The solid (red) and dashed (blue) arrows in redshift $3\le z \le 6$ indicate the halo masses
correspond to circular velocities $v_c = 35~$km~s$^{-1}$ and $95~$km~s$^{-1}$ respectively.
}
\label{fig_rates}
\end{figure*}
In Fig.~\ref{fig_rates} we show the ratio of 
the redshift derivatives for different halo mass functions to 
that obtained for the PS mass function.
In panels with $3\le z \le 6$ we show three characteristics masses related to
our star formation model at that redshift. The dotted vertical lines
show the mass that collapsed from $3\sigma $ fluctuations at that
redshift. In addition with solid (red) and dashed (blue)
arrows we show the masses corresponding
to halos with circular velocity $v_c = 35~km~s^{-1}$ and $95~km~s^{-1}$
respectively.
This is the mass range where radiative feedback from the metagalactic
UV background becomes important (Thoul \& Weinberg 1996, also see in Sec.~3).
In this redshift range the process
of reionization
is believed to be already over (Fan et al. 2006) and hence the
radiative feedback is well constrained in our model (indeed we will see
in the next section that all our models predict redshift of reionization
$z_{re}\ge 6$). We do not show these arrows for $z> 6$
as the exact radiative feedback will depend on the history of reionization. 
Thus, for $z> 6$ we only show the mass that collapsed from a $3\sigma $
fluctuations by dotted lines.

It is clear from the figure that for higher mass objects which are rare,
the PS mass function always predicts a smaller net halo formation rate compared
to any other mass function. For smaller mass objects the trend is
opposite. This will lead to different slope in the luminosity
function derived from different halo mass function.
At $z\ge 6$ all the mass functions
predict similar halo formation rate (within 10\%) around the
characteristic mass which corresponds to $3\sigma$ fluctuations.
However, at the typical halo masses  that will be detected as galaxies 
($\ge 10^{10} {\rm M}_\odot$)  PS underpredicts the net halo formation rate 
upto an order of magnitude.

For $z\lesssim 5$ the $3\sigma $ fluctuations are above $10^{12}~M_\odot$
and the predictions of other mass functions compared to
that of PS mass function are upto an order of magnitude higher. However,
in our modelling of star formation, we assume a sharp cut off in the
star formation for halos of mass above $10^{12}~M_\odot$, attributed to the
AGN feedback (see section~2 of Paper~I).
This reduces the difference in the prediction of our semi-analytical
modelling of star formation arising due to different halo mass 
functions at these redshifts.
Hence, for $z\lesssim 5$
all the mass functions predict a halo formation rate within a factor 2
of each other, for typical masses which contribute to the star formation.

The abrupt cut-off seen in the formation rate at $z = 3 $
 around $M= 10^8~M_\odot$
is due to the fact
the derivative of the PS mass function is becoming negative.
This shows the unphysical behavior in calculating the net formation
rate by taking derivative of the mass function.
However, star formation in such low mass halos ($M \lesssim 10^9~
M_\odot$) at $z = 3$ are suppressed by the radiative feedback (see below).
It is interesting to note that all the mass
functions other than the PS mass function predict similar ${\cal N}$.
This clearly shows that semi-analytical models using different mass
functions (except the PS mass function), will all give similar results.

It is a nontrivial exercise to demonstrate the difference between the formation
rate obtained from the Sasaki formalism of the PS mass function,
(${\cal N}_s (M,z,z_c)$),  to that calculated
by taking the derivative (${\cal N} (M, z_c)$). This is because in the 
Sasaki formalism one needs two redshifts, namely the collapse redshift 
($z_c$) of the halo  and the observed redshift ($z$) in order to calculate
the formation rate weighted by the survival probability. On the other hand 
the derivative contains only one redshift, the collapse redshift. 
However, we will try to show the difference between these two
approaches in the specific context of outflows in Section 5.

In the following sections we will show how the predictions of semi-analytic
galaxy formation models change with the mass function used.
We begin with a brief overview of our prescriptions
for star formation and reionization as in Paper~I and II.

\section{Star formation and Reionization} 
The star formation rate of an individual dark matter halo of mass $M$
collapsed at redshift $z_c$ and observed at redshift $z$ is modelled by
(Chiu \& Ostriker 2000),
\begin{eqnarray}
\dot{M}_{\rm SF}(M,z,z_c) &=& f_{*} \left(\f{\Omega_b}{\Omega_m} M \right) 
\f{t(z)-t(z_c)}{\kappa ^2~ t_{\rm dyn}^2(z_c)} 
\exp\left[-\f{t(z)-t(z_c)}{ \kappa ~t_{\rm dyn}(z_c)}\right].
\label{eqn_sfr}
\end{eqnarray}
Here, $f_*$ is the fraction of total baryonic mass that goes into stars
in the entire lifetime of the halo and $\kappa$ is a parameter
which governs the duration of the star formation activity in the halo.
As in Papers I and II we take $\kappa=1$ unless stated otherwise.
Further, $t(z)$ is the age of the universe at redshift $z$; thus
$t(z) - t(z_c)$ is the age of the galaxy and $t_{\rm dyn}$ is the
dynamical time at that epoch.
The star formation rate is converted to luminosity at $1500$~\AA~  
assuming an initial mass function (IMF) of the stars formed
[see Eq.~(6)-(8) of Paper~I].
Only a fraction ($1/\eta$) of this UV luminosity would come out
from the galaxy due to the absorption by dust.

The minimum critical mass of a halo ($M_{\rm low}$) which can sustain star formation
at a given epoch is decided by the cooling efficiency of the gas and also by 
radiative feedback in ionized regions. 
We consider models with  $M_{\rm low}$
corresponding to a virial temperature, $T_{\rm vir} = 10^4$~K ( as
``atomic cooled model'') and $300$~K (``molecular cooled model'').
For ionized regions of the universe,
our models assume complete
suppression of star formation in halos below a circular velocity
$v_c=35$~km~s$^{-1}$, no suppression above circular velocity of
$95~$km s$^{-1}$ and a linear fit  from $1$
to $0$ for the intermediate masses (as in Bromm \& Loeb 2002 ;
also see Benson et al. 2002; Dijkstra et al. 2004).
We have already indicated these characteristic masses in Fig.~\ref{fig_rates}
with solid (red) and dashed (blue) arrows.
The star formation in the high mass halos are also reduced by a suppression 
factor $[1+(M/10^{12} M_\odot)^3]^{-1}$ to incorporate possible AGN
type feedback in massive halos  as in Paper~I (also see Bower et al. 2005;
Best et al. 2006).
Given the luminosity evolution for an individual galaxy, and their abundance from
the halo mass function, one can then easily estimate
the UV luminosity function at any redshift (Eq.~(8) of Paper~I).

Ionization history of the universe is required to
self-consistently incorporate radiative feedback 
for each model.  We compute it following again the method
set out in Paper~I (section 2.4).
In Table~\ref{tab_reion} we summarize the reionization history in terms of the electron
scattering optical depth to the reionization ($\tau_e$) and
the redshift of reionization ($z_{re}$) for some of our models.
Note that $z_{re}$ is the redshift when the hydrogen ionization fraction becomes unity.
In all these calculations we have assumed only 10\% of the created 
UV photons 
escape into the IGM.
\begin{table}
\begin{center}
\caption{Reionization history}
\begin{tabular}{c c c c c}
\hline
Mass Function 	     & ${f_*^A}^\dag$ & ${f_*^M} ^\ddag$ &  $z_{re}$ & $\tau_e$ \\ \hline
\multicolumn{5}{c}{Atomic Cooled Models} \\ \hline
PS (Sasaki)	     & 0.30  & 	 - & 7.2      & 0.085    \\
PS (derivative)	     & 0.30  & 	 - & 7.2      & 0.085    \\
ST		     & 0.25  & 	 - & 7.6      & 0.096    \\
JW		     & 0.25  & 	 - & 7.8      & 0.096    \\
Warren               & 0.25  & 	 - & 7.6      & 0.095    \\
Reed                 & 0.25  & 	 - & 7.4      & 0.093    \\ \hline
\multicolumn{5}{c}{Molecular Cooled Models} \\
\hline
PS (Sasaki)	     & 0.30    & 0.10    & 6.2      & 0.120    \\ 
PS (Sasaki)	     & 0.30    & 0.05    & 5.9      & 0.105    \\ 
ST 		     & 0.25    & 0.10    & 6.5      & 0.134    \\ 
ST		     & 0.25    & 0.05    & 6.4      & 0.116    \\ \hline
\multicolumn{5}{l}{$^\dag$ $f_*^A$ is the $f_*$ in atomic cooled halos.}\\ 
\multicolumn{5}{l}{$^\ddag$ $f_*^M$ is the $f_*$ in molecular cooled halos.}\\
\end{tabular}
\label{tab_reion}
\end{center}
\end{table}
Table~\ref{tab_reion} 
shows that there is negligible difference
in $\tau_e$ and $z_{re}$ if one uses the derivative of the PS
mass function or the Sasaki formalism to calculate the net formation rate
of collapsed halos. 
For a range of $\kappa=0.1-4$, we find these differences
to remain within $4\%$. 
This also turns out to be the case for molecular cooled models.
Indeed the reionization histories are also almost identical.
However, the ST and other mass functions predict slightly
different reionization histories compared to the PS mass function 
though they do not differ greatly amongst themselves.
It is interesting that even if $f_*$
is larger for the model with the PS mass function compared
to that with other mass functions, 
$\tau_e$ is smaller in this model. 
{Thus to get the same $\tau_e$ we need a smaller efficiency
(or $f_*$) for models with ST and other mass functions
compared to models using the PS mass function.}
This basically reflects the smaller abundance
of rare high mass collapsed halos for the PS halo mass function at high $z$. 

However for `molecular cooled' models, one needs to
lower $f_*$  considerably in the molecular cooled objects, 
to be consistent with the observed value of $\tau_e$
for the ST and other mass functions.
This can be seen in Table~\ref{tab_reion}, that to
be consistent with the observed $\tau_e$, one needs to adopt
a value of $f_* < 0.05$ for molecular cooled halos.
Also note that $z_{re}$ is less
in molecular cooled models though $\tau_e$ is higher.
This is because of radiative feedback: more UV photons are generated by the
molecular cooled objects which increase the degree
of ionization of the IGM at a given redshift compared
to atomic cooled models, and causes early suppression
of star formation in halos with a circular velocity
less than $95~$km s$^{-1}$.
Hence it takes longer time to get complete
reionization of the IGM making $z_{re}$ less in molecular
cooled models.

Hence we conclude that the ST and other mass functions
leads to higher $\tau_e$ compared to the PS mass function
even when we use lower $f_*$.
Thus to reproduce observed $\tau_e$, the models using
other mass functions require less efficiency for star formation
and/or UV escape compared to that of models using the
PS mass function. 
If stars are formed
with `top-heavy' IMF in the molecular cooled low mass halos
then one has to further reduce the $f_*$ in those halos,  to
have a consistent $\tau_e$.
In particular the reionization efficiency of molecular
cooled halos has to be substantially reduced when one uses the
ST and other mass functions obtained from the simulation
instead of the PS mass function.

\section{UV Luminosity Functions}
\subsection{Comparison between different halo mass functions}

\begin{figure*}
\centerline{%
\includegraphics[bb=18 147 587 711,width=1.\textwidth]{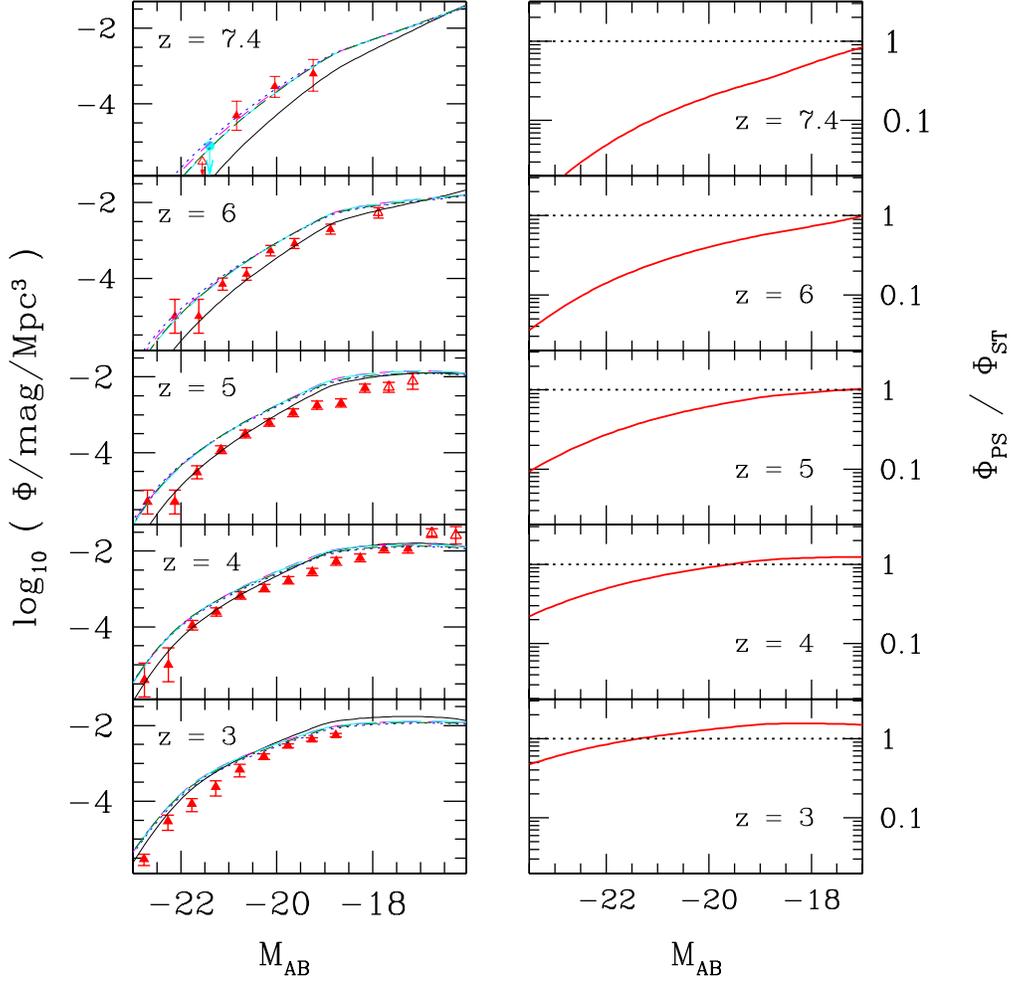}
}
\caption[]{Left panels: The UV luminosity functions of LBGs at different redshifts.
The observed data points are taken from
Reddy \& Steidel (2008) (triangles at $z=3$),
Bouwens et al. (2007) (triangles at $z = 4,~5$ \& $6$)
and Bouwens et al. (2008) (triangles at $ z = 7.4$).
The upper limit at $z=7.4$ is taken from Mannucci
et al. (2007) (the circle with an arrow).
The solid (black) lines are for the PS mass function with Sasaki
formalism. We have also shown the luminosity functions as predicted
from ST (dotted blue), JW (short dashed magenta), Warren (long dash
cyan) and Reed (dot short dash dark green) mass functions.
Right panels: We show the ratio of the luminosity functions
derived from the PS mass function to that from the
ST mass function.
}
\label{fig_UV_lf}
\end{figure*}

In this section we turn our attention to the UV luminosity function. First
we illustrate the differences between results from different mass functions for
fixed model parameters (see Fig.~\ref{fig_UV_lf}).
Here, we do not make any
attempt to fit the observed luminosity function. This will be done 
in the latter part of this section.

In our model we assume $f_* = 0.3$, $\kappa = 1.0$ with 
cosmological parameters mentioned in the introduction.
We apply a constant dust reddening correction of factor
$\eta = 4.5$ for all redshifts and masses.  
It is clear from this figure that at $z\gtrsim 4$, the PS
halo mass function (solid lines) predicts lower number of LBGs
compared to any other mass functions
in the observed luminosity range.
This is simply a manifestation
of the fact that the PS mass function predicts less number of rare high mass
halos compared to other mass functions. But at the low mass end
the PS mass function predicts a higher number density of collapsed
objects. This effect can be seen at $z=3$ where the UV luminosity
function at $M_{AB}\gtrsim -19$ is slightly higher for the
models with PS mass function.
The difference between the PS mass function and other mass functions
are mass dependent which leads to a different slope in the luminosity
function calculated from different mass functions.
As expected all the other halo mass functions
obtained from the simulation
predict almost the same luminosity function at all redshifts. 
It is interesting to see that all the mass functions predict almost
similar luminosity function at $z=3$ even though they are different
at higher redshifts. As we pointed out before 
the slopes of the luminosity functions
coming from different mass functions are indeed different.

In the right panels of Fig.~\ref{fig_UV_lf} we show the ratio
of luminosity function obtained from the PS mass function
to that from the ST mass function.
It is clear from the figure that there is a characteristics
luminosity (or in turn the halo mass) for each redshift
where the amplitude of the luminosity function derived
from the PS and ST mass functions match with each other.
However this characteristic luminosity increases with
decreasing redshift.
At $z\sim 3$, in the mass
range of halos where star formation takes place the PS mass function
over produces halos compared to the ST mass function
(see Fig.~\ref{fig_rates}).
 At the low mass
range the excess is roughly a factor 2. The excess seen at the
low end of the luminosity function
produced by the PS mass function is due to this.
At $M\sim 10^{12}~M_\odot$ all the models produce same 
number density at $z=3$. As star formation in halos above this mass are
suppressed we do not see order of magnitude deviation in the high luminosity
end as we would have expected from Fig.~\ref{fig_rates}.
However at $z \sim 5$, compared to the PS mass function the ST mass function
produces a factor $\sim 2$ excess at $M = 10^{12}~M_\odot$ and similar number
of halos at the mass corresponding to $v_c=95~$km~s$^{-1}$. Thus
we see a much better matching at the low luminosity end of the luminosity
function (see Fig.~\ref{fig_UV_lf}) and a very strong deviation
at the high luminosity end
for $z\gtrsim 5$.

Thus the reasonable match in the luminosity function at $z=3$
 and strong disagreement at high-$z$ could be attributed to the
fact that as one moves to higher and higher redshifts
a given luminosity range is contributed
by rarer and rarer objects (or higher $\sigma$ fluctuations;
see Fig.~\ref{fig_rates}) and the
PS mass function is known to predict less number of rare objects 
compared to the ST mass function or other mass functions
derived from simulations.

We find that the luminosity functions
calculated using the derivative of the PS mass function
and that obtained from the Sasaki formalism, differ negligibly
for $\kappa=1$. However for $\kappa=0.1$
the Sasaki formalism predicts larger abundances by about
$30\%$, in the observed luminosity range. The
reason is as follows: first, the Sasaki formalism for 
$z=z_c$ gives the formation rate
of halos, while the derivative gives the formation minus the
destruction rate, and hence a lower net formation rate.
Second, for any $\kappa$ we detect galaxies whose ages are
$t_a \sim \kappa t_{\rm dyn}$, when they shine the maximum. 
In the Sasaki formalism,
only a fraction of halos $p=D(t-t_a)/D(t)$ formed at $(t-t_a)$ survive 
at the observed time $t$.
For small $\kappa$, we have $t_a/t \ll 1$, and thus $p\sim 1$
and the luminosity function predicted by the 
Sasaki formalism is higher, as its formation rate is
higher. However as $\kappa$ increases,
$t_a$ increases, $p$ decreases, and hence luminosity functions computed by
the two methods first approach each other for $\kappa \sim 1$,
after which the derivative method starts to produce
a larger luminosity function. Nevertheless, since $t_{\rm dyn}$ 
is much smaller than the Hubble time, we find the differences 
between the luminosity function calculated
by these two methods, are less than $30\%$ for all $z$.
This difference is much smaller than the difference 
between luminosity functions, that are obtained using
PS and other mass functions for $z>3$.

\begin{figure*}
\centerline{%
\includegraphics[bb=18 147 587 711,width=1.\textwidth]{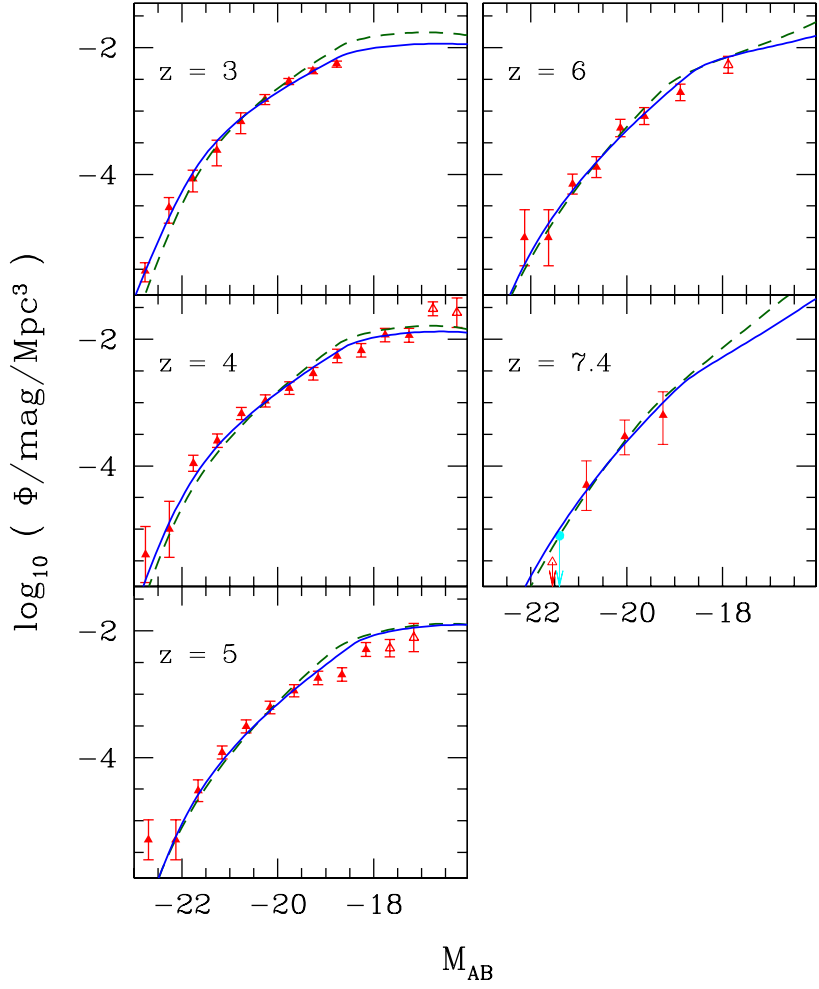}
}
\caption[]{UV Luminosity functions of LBGs at different redshifts for the
ST (solid lines)  and PS (dashed lines)
mass function. We fit the observed
data points by changing $f_*/\eta$ using a $\chi^2$-fit.
We ignore few data points in the low luminosity end (open
triangles) while fitting as they suffer
from completeness problem.
}
\label{fig_UV_lf_fit}
\end{figure*}

It is also clear from Fig.~\ref{fig_UV_lf} that the shape of the
UV luminosity functions (the slope and characteristic
luminosity)  and its redshift evolution, does depend on the assumed form
of the halo mass function. 
This is expected as we have seen in Fig.~\ref{fig_rates} that the difference
of halo formation
rate coming out from different mass functions are mass dependent.
Hence it suggests that, in principle
we would be able to constrain the form of the halo mass function,
just by fitting the luminosity functions. In what follows we use
$\chi^2$ minimization to address this issue.

\subsection{Best fit models: $\chi^2$ minimization}

We use $\chi^2$ minimization technique to quantify how well the models 
using different mass functions reproduce the observed luminosity functions.
For each redshift bin we have used the most recent measurement of 
the luminosity function that covers a wide range in luminosity. It is known
that the estimations of luminosity function of LBGs suffer from the lack of
redshift information and cosmic variance much more than typical Poisson
errors (for example see Beckwith et al. 2006; Reddy et al. 2008; 
Bouwens et al. 2007; 2009). As $\chi^2$ minimization technique is sensitive
to the errors, in our analysis we use errors that take care of the
cosmic variance in addition to the Poisson error. We consider $f_*/\eta$ as the 
only free parameter. As this is assumed to be constant over the whole 
mass range, varying $f_*/\eta$ is like changing the mass to light ratio.
Note that the observed luminosity functions are also given at different rest wavelength.
For example at $z=$~3, 4, 5, 6 and 7.4 the observed data points are
obtained at $\lambda =$~1700~\AA, 1600~\AA, 1600~\AA, 1350~\AA~
and 1900~\AA~ respectively.  
In our models we take care of this aspect by calculating 
the intrinsic luminosity at appropriate wavelengths.
The observed data and the best fit LF for the ST (solid line) and
PS (dashed line)
mass functions are given Fig.~\ref{fig_UV_lf_fit}. The best fit
parameters are summarized in Table~\ref{tab_fit}.

For $z\sim3$ we use the observed luminosity function given by
Reddy \& Steidel (2008) which covers the low luminosity end
well. As this LF is obtained using several independent fields,
the errors take into account the cosmic variance very well.
As can be seen from the Table~\ref{tab_fit} the model
using the ST mass function provides a good fit to the data
(with a reduced $\chi^2_\nu = 0.93$). 
Whereas the reduced $\chi^2$ for the model using the PS mass function is
high ($\chi^2_\nu = 3.90$). This is mainly because the PS
mass function overpredicts low luminosity
objects and under predicts high luminosity objects.  This 
together with the incompleteness in the observational data
used in Paper~I prompted us to say in Paper~I that
one needs additional feedback
to suppress the star formation in low mass halos that
contribute to LF at the low luminosity end.
However, it is clear now that for models with ST mass function,
no additional feedback is needed to fit the luminosity function
at $z\sim3$. We also tried to fit the data by changing the 
characteristic mass scale for the AGN feedback. Even then
the reduced $\chi^2$ for the best fitted models with the PS mass function
are always larger than 2.

\begin{table}[h]
\begin{center}
\caption[The best fit values of $f_*/\eta$ for different redshifts]
{The best fit values of $f_*/\eta$ for different redshifts.
 We show best fit parameters and $\chi^2$
per degree of freedom for models with both ST and PS halo mass functions.
}
\begin{tabular}{c | c | c| c| c| c | c}
\multicolumn{7}{c}{}\\
\hline \hline
& \multicolumn{3}{|c|}{ST}& \multicolumn{3}{c}{PS} \\ \cline{2-7}
{\raisebox{1.5ex}[0cm][0cm]{$ z $}} & $f_*/\eta^\dagger $ & $ \chi^2_\nu$ & $(f_*/\eta)_{1500}^\ddagger $ 
& $f_*/\eta ^\dagger $ & $ \chi^2_\nu$ & $(f_*/\eta)_{1500}^\ddagger $ \\ \hline
$3$ & $0.066 \pm 0.001$ & $0.97$ & $0.055$ & $0.053 \pm 0.001$ & $3.90$ & $0.044$  \\ \hline
$4$ & $0.049 \pm 0.001$ & $1.09$ & $0.042$ & $0.053 \pm 0.001$ & $2.32$ & $0.046$ \\ \hline
$5$ & $0.040 \pm 0.001$ & $4.85$ & $0.034$ & $0.049 \pm 0.001$ & $8.28$ & $0.042$ \\ \hline
$6$ & $0.036 \pm 0.001$ & $0.63$ & $0.050$ & $0.058 \pm 0.002$ & $1.19$ & $0.081$ \\ \hline
$7.4$ & $0.111 \pm 0.005$ & $1.42$ & $0.100$ & $0.207\pm 0.009$ & $2.27$ & $0.186$ \\ \hline \hline
\multicolumn{7}{c}{}\\
\multicolumn{7}{l}{$^\dagger$ $f_*/\eta$ obtained at the observed wavelength}\\
\multicolumn{7}{l}{$^\ddagger$ $f_*/\eta$ calculated at $\lambda=1500~$\AA~
considering a dust model similar to} \\
\multicolumn{7}{l}{~~~that of our Galaxy or large Magellanic clouds \citep{dustmodel}}\\
\end{tabular}

\label{tab_fit}
\end{center}
\end{table}

We now consider the luminosity functions for $z\sim 4$.
The observed luminosity function at this redshift covers a much wider
luminosity range, thanks to the Hubble Ultra deep field (HUDF) and GOODS
(The Great Observatories Origins Deep Survey)
data. Tabulated luminosity functions based on HUDF and GOODS are available
in Beckwith et al. (2006) and Bouwens et al. (2007). We use Bouwens et al's
data for our analysis. They estimate rms due to cosmic variance to be 
$\sim 22$\%. This fractional uncertainty is added to the Poisson error
in quadrature (as done in Beckwith et al. 2006 and Bouwens et al 2009). 
We find that this is essential to get $\chi^2_\nu \sim 1$ while fitting 
even their best fitted  Schechter function to the UV LF given in Table~5 
of Bouwens et al. (2007).
Also we ignore the last two data points at the low luminosity end
(shown by the open triangles in Fig.~\ref{fig_UV_lf_fit})
while fitting as these points are affected by incompleteness
of the survey (Bouwens et al., 2007). Even for this redshift bin the model using
the ST mass function
provides better fit to the data than the one that uses the PS mass function.
It is also interesting to note that $f_*/\eta$ is nearly constant for
the model that uses the
PS mass function between $z\sim3$ and $z\sim4$ but decreases for models using ST
mass function.

The observed luminosity functions for $z\sim5$ using HUDF
are available in Oesch et al. (2007) and Bouwens et al. (2007). 
We use Bouwens et al's data for our analysis. A fractional uncertainty of
$\sim 18\%$ (Bowens et al. 2007) is added to the Poisson error in quadrature.
We ignore the last two data points at the low luminosity end
due to incompleteness (Bouwens et al., 2007).
It is clear from Table~\ref{tab_fit} and Fig.~\ref{fig_UV_lf_fit} 
that the observed LF at this redshift is not reproduced well either by
models using ST mass function or PS mass function. Observed points 
in the lower luminosity end are systematically lower than model
predictions. In particular the discontinuity seen at $M_{AB} = -18.6$ 
contribute appreciably to the $\chi^2$. Note such a structure in
LF, if true,  can not be reproduced by our models.
However, what is interesting is that
for this redshift bin the model using the ST mass function
provides better fit to the data than the one that uses the PS mass function.

The observed luminosity function at $z\sim 6$ are from Bouwens et al. (2007).
 A fractional uncertainty of
$\sim 22\%$ (Bowens et al. 2007)  is added to the Poisson error in quadrature. 
Models with both PS and ST mass functions provide acceptable fits to the data,
with the latter having better reduced $\chi^2$. The observed luminosity
function at $z\sim 7$ are from (Bouwens et al. 2009) and the published errors 
already account for the uncertainty due to small volume sampled.  As the
number of constraints are less models with both the mass function 
reproduce the data well.

The interesting result that emerges from the analysis presented till now is that
the models with the ST mass function reproduce the observed LF much better
than the ones that use the PS mass function. We also checked whether this
result is valid if
we use different observed luminosity functions. We find our best fit value of
$f_*/\eta$ depends very much on the observed luminosity function we use
(similar to what one finds for the Schechter function parameters). 
For example, for $4\le z\le 6$ the observed luminosity functions
lack high precision measurements in the high luminosity end due to the small
volume probed. Thus our best fit $f_*/\eta$ is decided by how well 
we fit the low luminosity end of the luminosity functions. However, ground 
based measurements which sample the high luminosity end well, tend to have 
slightly higher $f_*/\eta$. Ideally one should merge the ground and 
space based measurements by carefully accounting for all the possible 
biases. This is a non-trivial exercise and beyond the scope of
this present work.
However, irrespective of the data we used, we find invariably that
the models with the ST mass function provide a better fit to the observed
data. At least for $z\sim 3$, where the LF is well defined in the high 
luminosity end, we find that even changing the AGN feed back in model
that uses PS mass function does not help in producing a $\chi^2$ better than
that obtained for models with ST mass function.

We now consider the best fit values of $f_*/\eta$ at different redshifts.
Columns~2 and 5 of Table~\ref{tab_fit} list
the best fit values of $f_*/\eta$ obtained at the observed wavelength
for which the luminosity function has been measured, for the models
with the ST and PS mass functions respectively. Note that in principle,
the amount of dust reddening correction
depends very much on the wavelength. Hence, in order to compare
the best fit values of $f_*/\eta$ at different redshifts we
use the dust model like that applicable to the Galaxy or large Magellanic
clouds \citep{dustmodel},
and bring all the $f_*/\eta$ values to the wavelength of $1500$~\AA.
These values are also tabulated in
Table~\ref{tab_fit} (in Column~4 and 7 for the
models with the ST and PS mass functions respectively).
It is clear from the table that no particular trend emerges
at $3\le z \le 5$ though for higher redshifts an increase
of $f_*/\eta$ may be needed. As already mentioned, the best fit values
of $f_*/\eta$ very much depend on the observed luminosity
function we have used. Also the nature of dust in the high redshift
galaxies is very uncertain. Hence at this stage it is difficult
to predict any trend on the values of $f_*/\eta$ and hence
on the evolution of star formation in the high redshift galaxies.
Future improved luminosity function measurements with reduced errors are
needed to probe the redshift evolution of $f_*/\eta$.
Nevertheless, as we noted above, it is heartening to find that
a fairly simple model incorporating some of the physically motivated
feedback effects and one free parameter $f_*/\eta$ does allow us
to explain reasonably the observed UV luminosity function
of high redshift LBGs.

Note that in Paper~I we showed that the
inclusion of star formation in the molecular cooled objects does not affect
the UV luminosity function in the redshift range considered here. Hence
we do not show any results of UV luminosity
function for molecular cooled models. However, the molecular
cooled objects play an important role in setting up the metallicity floor
in the IGM at high redshift as shown in Paper~II. Therefore, we will
also consider the models with star formation in molecular cooled objects
for galactic outflows.

\section{Galactic outflows and the IGM}

We turn now to models of galactic outflows and their feedback into the
IGM, an issue that was addressed in detail in Paper II. 
Note that the metals detected in the IGM can only have been synthesized by stars
in galaxies, and galactic outflows are the primary means by which they
can be transported from galaxies into the IGM 
{
(Silk, Wyse \& Shields, 1987, Tegmark, Silk \& Evrad 1993, Miralda-Escude \& Rees, 1997,
Nath \& Trentham, 1997, 
Madau, Ferrara, Rees 2001, Furlanetto \& Loeb 2003,
Scannapieco 2005, Bertone, Stoehr \& White 2005,
 Oppenheimer \& Dave 2006, Bertone, De Lucia \& Thomas 2007).
}
The mechanical energy
that drives outflows arises from the supernovae (SNe) explosions
associated with the star formation activities in the galaxy.
We concentrate
on how much of the IGM can be polluted by the outflows from
galaxies and also what would be the metallicity of those
polluted regions. We follow Paper~II in modelling
galactic outflows and their feedback effects. Here we briefly outline 
this procedure.

First, the star formation rate of an individual galaxy is converted to the
rate of SNe. This depends on the assumed IMF. 
For the IMF used in this work, one SNe is produced
per 50 solar mass of star formation. A single SNe is assumed to 
produce $10^{51}$~erg of energy out of which a fraction 
$\epsilon_w=0.1$ goes into the outflow.
The outflow is taken to be spherically symmetric and
its dynamics is followed using a 
thin-shell approximation (see Eqs.~(5)-(16) of Paper~II for the details).
The dark matter potential of the halo is assumed to be in NFW profile
and a fraction ($f_h = 0.10$) of the total baryonic mass is taken to be
in thermal equilibrium at virial temperature in this potential. The amount of mass
coming out of the star forming region is assumed to be proportional
to the mass of stars formed
with the proportionality constant $\eta_w = 0.3$.
We also assume that $90\%$ of the shocked IGM/halo gas is concentrated
in the thin-shell region and rest $10\%$ is incorporated in the
hot bubble. The initial radius of the shell is taken to be $1/10$ the virial
radius, and the outflow is frozen to the Hubble flow when its peculiar velocity
drops to the sound speed of the surrounding IGM.

The extent $R_s$, to which an individual outflow can propagate into the IGM, 
depends mainly on the rate of SNe productions and the energy efficiency 
of the outflow, $\epsilon_w$. In Paper~II, we showed that
$R_s$ is fairly independent of other parameters 
as long as the outflow escapes the halo potential.
Note that we have already constrained $f_*/\eta$ by
fitting the UV luminosity functions of LBGs. However,
we need some specific values of $f_*$ to calculate the effect
of outflow. For the PS mass function we take
$f_* = 0.30$ and for ST and other mass
functions $f_* = 0.25$ for the whole $z$ range.

After calculating the evolution of a suite of individual outflow
models, we can study several global properties of the wind affected
regions. One simple quantity is the porosity $Q(z)$,
which is a measure of the (volume) fraction of the universe affected by outflows. 
This is calculated by adding up the outflow volumes around all the sources at any redshift:
\begin{equation}
Q(z) = \int \limits_{M_{\rm low}}^{\infty} \de M \int \limits_z ^{\infty}
\de z_c  ~{\cal N} (M, z_c)~ \f{4}{3}\pi \left[R_S(1+z)\right]^3.
\label{eqn_q}
\end{equation} 
Here, $R_s(M,z,z_c)$ is the radius of an outflow at redshift $z$, arising from a halo
of mass $M$ which collapsed at redshift $z_c$. This  
comes from solving for the outflow dynamics.
Note that in the Sasaki formalism as applied to the PS mass function, 
one has to replace ${\cal N} (M, z_c)$ by  ${\cal N}_s (M,z, z_c)$ 
(see Paper~II). 
For $Q < 1$ the porosity gives the probability that a randomly selected point 
in the universe at $z$ lies within an outflow region. 
For $Q \sim 1$, it is more useful to
define the associated filling factor of the outflow regions,
which if outflows are randomly distributed is given by, 
$F=1-\exp [-Q(z)]$.
One can also calculate porosity weighted averages of various physical quantities associated
with the outflows, and their probability distribution functions (PDFs). 
In our models, we compute the metallicity of the outflowing
gas and the global average metallicity assuming instantaneous metal
mixing within the galaxy (see appendix~A of Paper~II).

\begin{figure}
\centerline{%
\includegraphics[width=18cm]{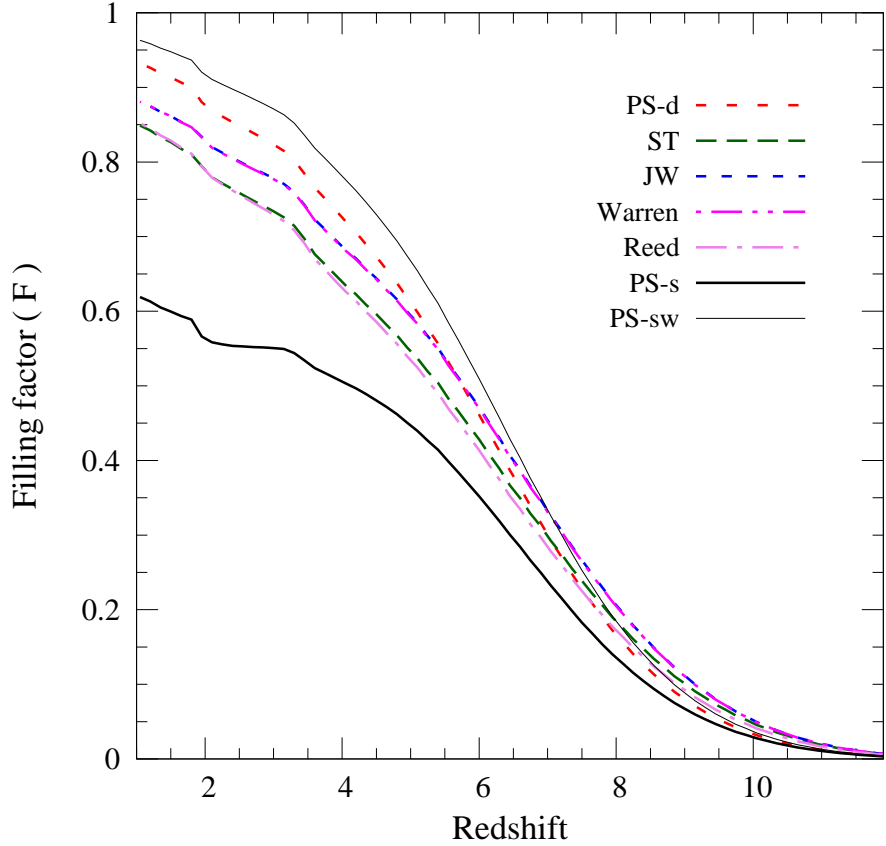}
%}
}
\caption[]{The volume filling factor of the IGM by the outflows
for models with
different halo mass functions. The dotted (red), long dashed
(green), short dashed (blue), dot dot dashed (magenta) and
dotted dashed(violet) curves are for the model
with the derivative of the PS, ST, JW, Warren and Reed mass function
respectively.
We also show the filling factor calculated using the Sasaki formalism
of the PS mass function with thick solid (black) line. {The thin solid (black)
line is obtained assuming no destruction of outflows, and the formation rate of halos
from the Sasaki formalism applied to the PS mass function.}
}
\label{fig_filling}
\end{figure}

In Fig.~\ref{fig_filling} we show the filling factor of the universe
as predicted by using different halo mass functions.
The different models considered are: the ST (long dashed
green), JW (short dashed blue), Warren (dot dot dashed magenta) and
Reed (dotted dashed violet) halo mass functions. 
In all these
models we assume $f_*=0.25$. For comparison we also show
the filling factor as calculated from the the PS mass function
using both the Sasaki formalism (thick solid black line) and
by taking the derivative to compute the net formation rate (the short dashed
red line). 
Recall that in these models
we take $f_*=0.30$.
It is clear that all the models which employ the derivative ${\cal N}$ 
to characterise the net halo formation rate,
predict similar results for the volume filling factor (with in 10\%),
with $F$ of order $0.7-0.8$ at $z=3$.
However the model using the Sasaki formalism and the PS mass function 
predicts a lower volume filling factor ($F = 0.55$ at $z=3$).
At $z\gtrsim 6$ the difference in $F$ is like 30\% where as at $z\lesssim 3$
the difference is about 50\%. We find as in Paper~II that outflows from halos
in the mass range $M\sim 10^7-10^9~M_\odot$ dominate the volume filling factor.
 
It is interesting that for the PS mass function, the volume filling factor 
is different for the two ways
of calculating the net formation rate, 
namely the derivative and the Sasaki formalism. 
\begin{figure}
\centerline{%
\includegraphics[width=21cm]{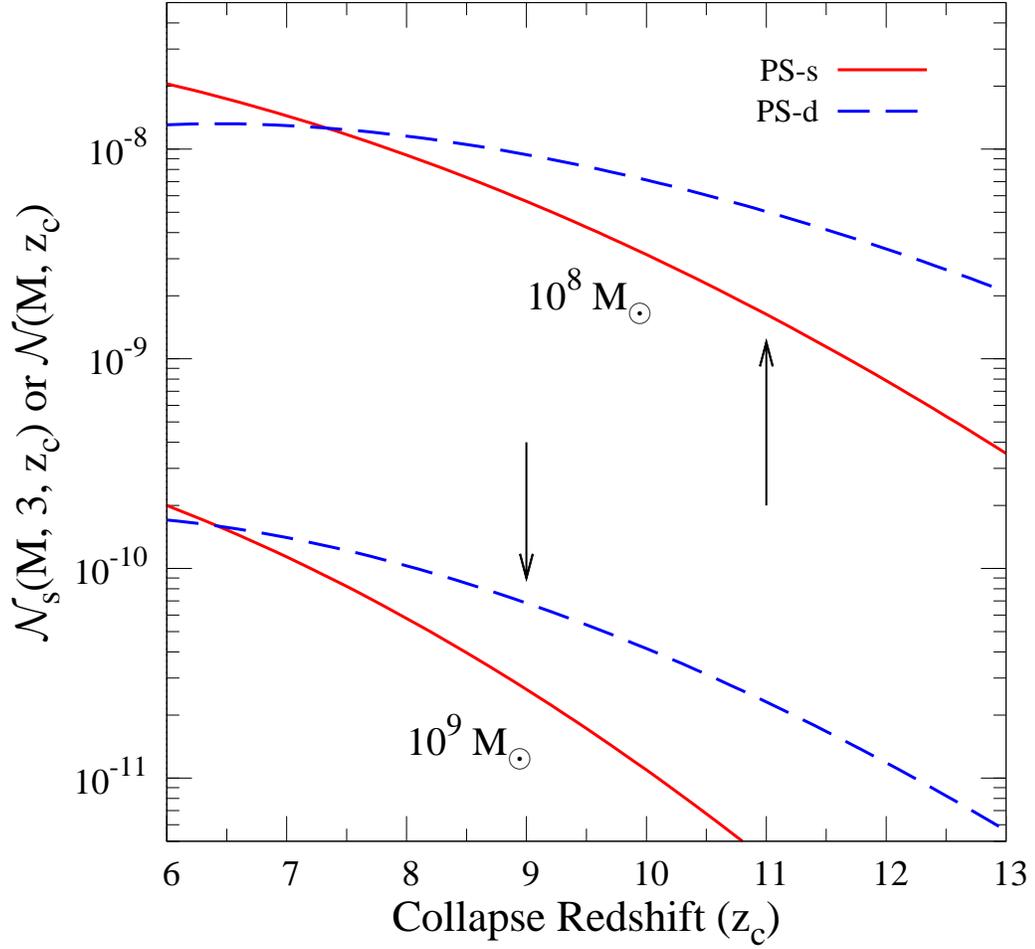}
}
\caption[]{Comparison of formation rate of the dark matter halos as obtained
from Sasaki formalism of the PS mass function (solid line) and that calculated
by taking the derivative (dashed line) for $10^8~M_\odot$ and  $10^9~M_\odot$
halos. The rates are in arbitrary units.
We fixed the observed redshift $z = 3$.
The arrows indicate the typical collapsed redshifts for these halos.}
\label{fig_sasaki_deriv}
\end{figure}
{This obtains although there was hardly any difference in the
UV luminosity function calculated from these two methods. The reason behind this
can be understood as follows.
In case of Sasaki formalism
whenever a halo is destroyed, one implicitly assumes that the outflowing
material 
no longer contributes to the volume filling factor.}
But in the models with the derivative of the PS mass function
once an outflow is formed it contributes to the volume filling
factor for ever ; although the derivative itself gives the net (formation 
minus destruction) rate, at $z_c$.
{And thus leads to a smaller formation rate of halos at $z_c$.}
 The net difference between Sasaki
formalism and the derivative of the PS mass function can be
seen from Fig.~\ref{fig_sasaki_deriv}. 

In this figure we show ${\cal N}_s (M,z,z_c)$
and ${\cal N} (M,z_c)$  as a function of $z_c$ for  $10^8~M_\odot$
and  $10^9~M_\odot$ halos, while fixing $z = 3.0$.
Note  that ${\cal N}_s (M,z,z_c)$
and ${\cal N} (M,z_c)$ come as a part of the integrand over $z_c$
while calculating the porosity of
the galactic outflows at a given $z$ (Eq.~\ref{eqn_q}).
We choose these values of $M$ as outflowing material from halos
with such masses dominate
the volume filling factor of IGM, as mentioned above.
It is clear from the
figure that at typical collapsed
redshifts for such halos (as indicated with the arrows) the derivative of the PS mass functions
predicts about a factor of $3$ higher net formation rate compared to that of
Sasaki formalism applied to the PS mass function. This difference decreases with 
decreasing $z_c$ and at certain redshift the Sasaki formalism predicts higher
formation rate {(weighted by the destruction probability)}
compared to the derivative. However
the star formation in halos with these masses would be suppressed by
the radiation feedback due to reionization at $z_c<$6.
Therefore, at $z=3$, the integrand in Eq.~\ref{eqn_q} will be larger
for the PS derivative compared to that obtained from Sasaki formalism
(for the range of halo masses and $z_c$ that contributes dominantly to $F$).
 For this reason the model with the
derivative of the PS mass function predicts
higher volume filling factor even if the UV luminosity function
predicted by them were same.

{
Note that when we use the Sasaki formalism, we could
go to the other extreme and not destroy any
outflow, by replacing the survival probability, $D(z)/D(z_c)$,
with unity. In this case we find that the volume filling factor
is the largest, even higher than that obtained when one uses
the derivative of the mass function to calculate the halo
formation rate.
We show this as a thin solid line in Fig.~\ref{fig_filling}.
The volume filling factor that realistically obtains
will be bracketed by these two cases (the thick/thin
solid lines of Fig.~\ref{fig_filling}).
This difference also gives an estimate of the uncertainty
in the volume filling factor obtained in semi-analytic models.
}

We see from Fig~\ref{fig_rates} that the
net formation rates, calculated using the derivative of all
the mass functions, agree reasonably well, for halos
that contributes dominantly to $F$. 
This is the reason behind the close agreement behind $F$ estimated
using the derivative of different mass functions.

Next we compare various physical properties of outflows
for different halo mass functions. 
\begin{figure}
\centerline{
\includegraphics[bb=40 60 565 530,width=15cm,height=13cm]{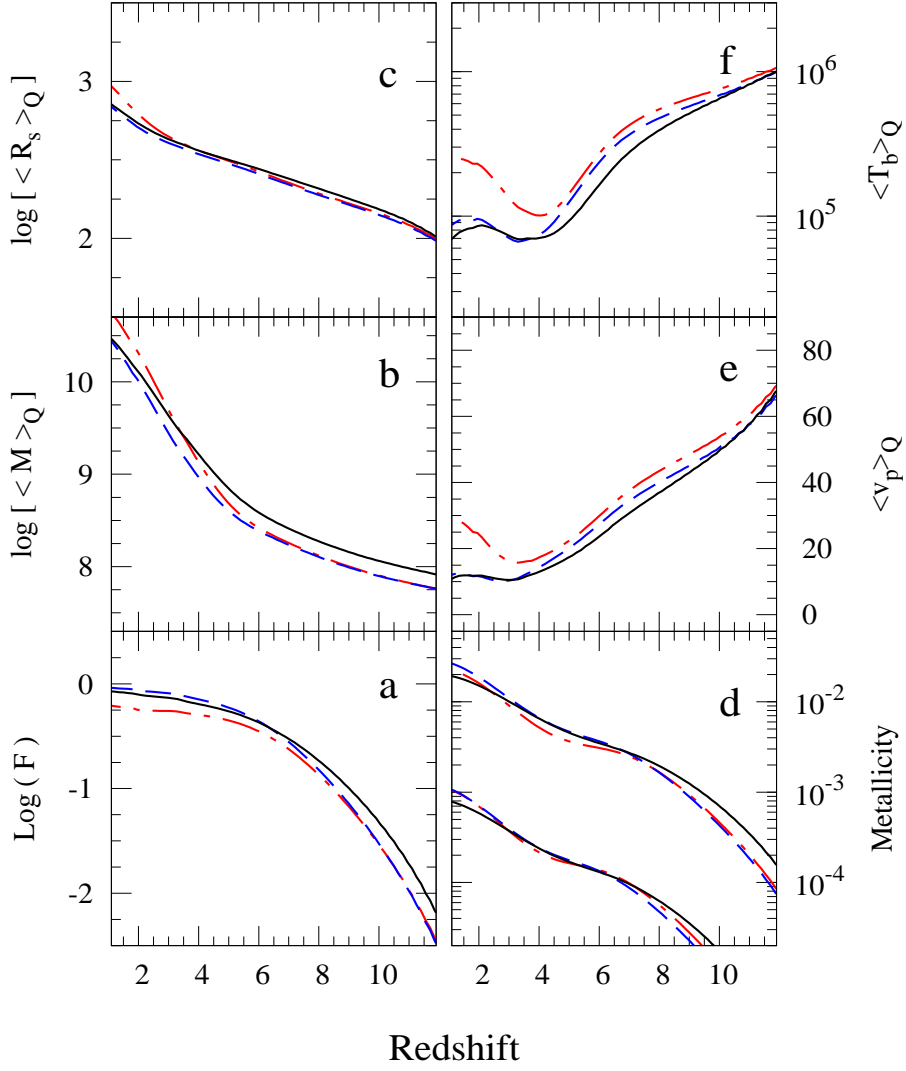}
}
\caption[]{The global properties of outflows for the derivative (dashes)
and Sasaki formalism (dotted dashes) of the PS mass function. We also
show for the derivative of the ST mass function with solid lines.
Panel (a) shows the volume filling factor $F$.
In panel (b) and (c) we show the porosity weighted average
dark matter mass (in $M_\odot$) and comoving radius (in kpc)
respectively. In panel (d) we show the global average metallicity
produced by star formation (the top set of curves), and that injected into the
IGM (the lower set of curves).
The porosity weighted average peculiar velocity (in km~s$^{-1}$) and
hot bubble temperature of the outflows
are shown in panel (e) and (f) respectively.
}
\label{fig_wind_avg}
\end{figure}
In Fig.~\ref{fig_wind_avg} we show various physical parameters
related to outflows for the derivative of the PS (dashed)
and ST (solid) mass function. For comparison we also show
the results for the Sasaki formalism of the PS mass function
with dotted dashed line.
 In panel (a) we compare the filling
factor that we have already discussed in great details.
 Panel (b) compares
the porosity weighted average mass of the dark matter halos that
contribute to the filling factor. Since the ST mass function
predicts more high mass objects, the average halo
mass that contribute to the filling factor predicted by this model
is slightly higher compared to the PS mass function
for $z\gtrsim 3$. However, vary massive halos which formed at lower redshifts
do not have an outflow due to smaller value of $f_*$. Hence at $z\lesssim 3$
average mass contributed to the volume filling factor is less in the models
with the ST mass function compared to that with the PS mass function.
Panel (c) compares
the porosity weighted average radius of outflows 
whereas panel (d) compares the global average metallicity 
produced by star formation (the top set of curves), and
that which is injected into the IGM (the lower set of curves).
Due to the reason already mentioned above, the average radius of the
outflow and the metallicity show a similar behavior as
the average halo mass for different models. In panel (e) and panel (f)
we show the porosity averaged peculiar velocity of the outflow
and the temperature of the hot bubble respectively.
Both the average peculiar velocity and the temperature
are higher in case of the Sasaki formalism mainly reflecting the fact that
$f_*$ is higher for the PS mass function compared to the ST mass function.
For the PS mass function, although the Sasaki formalism 
and the derivative have the same $f_*$
the average peculiar velocity and the temperature
are lower in the model with the derivative due to the
higher volume filling factor. In general, we conclude
from Fig.~\ref{fig_wind_avg} that even if the filling factor changes
depending on the way we calculate the
net formation rate (upto 50\% change), the physical properties related to the outflow do not
vary significantly between these models.
As mentioned earlier, an important invariant feature of all the outflow models
is that small mass halos with $M \sim 10^7-10^9 M_\odot$ contribute dominantly
to the porosity $Q(z)$ {(as originally discussed by Silk, Wyse \& Shields, 1987,
Miralda-Escude \& Rees, 1997, Nath \& Trentham, 1997, Madau, Ferrara \& Rees, 2001).}

{
The mildly overdense regions which produce the Lyman-$\alpha$
forest lines probe the power spectrum of density perturbations
down to a comoving scale of half a Mpc or so
at $z\sim 3$. 
In Paper~II we found that in the atomic cooled models,
the IGM is polluted by regions of hot gas ($T\sim 10^5$~K)
with sizes of $200-800$~kpc.
This could potentially distort the density power spectrum as
inferred from the Lyman-$\alpha$ forest.
There is no evidence
in the current data for such distortions. However, this aspect
needs to be studied further to see how much of a problem this is
for the above models.
}

On the other hand, we also noted that molecular cooled models
are better placed to seed the IGM with metals without strong 
distorting 
effects on the Lyman-$\alpha$ forest lines. 
This is because outflows from molecular cooled halos 
can significantly fill the universe, even at $z\sim 8$, while at the same time
having typically a much smaller radius.
We therefore also consider here
the effect of changing the halo mass function in such models.
\begin{figure}
\centerline{
\includegraphics[width=17cm]{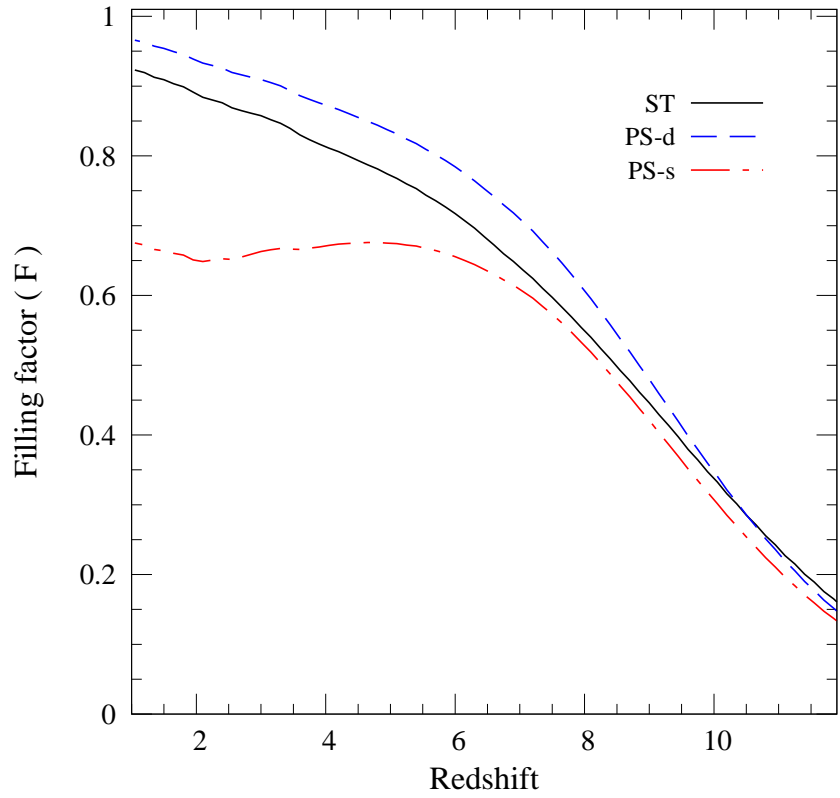}
%}
}
\caption[]{Volume filling factor as obtained from molecular cooled models.
The solid and dashed lines are for the models with derivative of the
ST and PS mass function. The dotted-dashed line show the volume filling
factor as obtained from Sasaki formalism of the PS mass function.
}
\label{fig_wind_m}
\end{figure}

In Fig.~\ref{fig_wind_m} we show the volume filling factor
predicted by the molecular cooled models,
as obtained from the models with derivative
of the PS (dashed) and ST (solid) mass function. We also
show the $F(z)$ when we use the Sasaki formalism and the PS mass function
(dotted dashed line).
{
We take $f_*=0.05$ in the molecular
cooled halos in order to be consistent with the optical
depth of reionization inferred from the CMB polarization measurements
(see Table~\ref{tab_reion}).
As expected all these models fill the IGM with outflows and metals
at fairly high redshift ($z\gtrsim 8$).
}
The volume filling factor is again smaller when one employs
the Sasaki formalism, essentially for the same reason as discussed in the
case of atomic cooled models. 
The higher value of $f_*$ in the model with the derivative
of the PS mass function compared to the model with the
derivative of the ST mass function makes the volume filling factor
larger for the former model.
We have also
checked that physical parameters in these models are similar
to that obtained in the molecular cooled models discussed in 
Paper~II. Hence the conclusions
about the molecular cooled models arrived at in Paper~II
remain; that they are more favorable in spreading the metals,
without unduly distorting the Lyman-$\alpha$ forest.

\section{Discussions and Conclusions}

In a series of papers (Paper I and Paper II) we have been developing
semi-analytic models of star formation and associated outflows 
in high redshift galaxies 
that are constrained by available observations on
UV luminosity function and reionization. Our models successfully explained
the observed UV luminosity functions of Lyman break galaxies 
at $z\ge 3$ and predicted various effects of outflows form 
the galaxies on the metal enrichment of the IGM. In all these
models one of the main ingredients is the halo mass function.
We used the standard PS mass function with the halo
formation rate given by Sasaki (1994). 
However the halo mass functions determined by fitting various 
numerical simulations of galaxy formation differ considerably
from the PS mass function, especially for rare objects.
It is then necessary to examine how this affects the 
model predictions given in Paper I and II.
Here, we present a systematic comparison of the results for different 
analytical forms of the halo mass functions.

Further, two methods are normally used in the literature
to compute the net formation rate of halos for the PS mass function; 
(i) the formalism of Sasaki is used to calculate a formation rate of halos at collapse
and then fold in the halo survival probability to later epochs,  
or (ii) the derivative
of the PS mass function is employed to get a net formation rate 
(formation rate minus destruction rate) at any
redshift. We had employed the Sasaki formalism for the PS mass function in
Paper I and II. 
Here we also calculate model predictions using the 
derivative of the PS mass function to test the
sensitivity of the results to the manner of
computing the net halo formation rate.
Such a comparison is also important as the Sasaki
formalism is not easily generalizable to
other mass functions. For all other halo mass functions
one has to simply take recourse to its 
derivative, to model the net formation rate of halos.

We first show, by comparing the derivatives of all mass functions,
that the historically used Press-Schechter
mass function predicts a lower formation rate of rare high
mass dark matter halos
compared to any other mass function. 
However, all other mass functions do not
differ significantly in 
the formation rate of collapsed dark matter halos.
Therefore, significant differences can arise in model predictions
made using the PS mass function and
the other forms of the mass functions considered here. 
We show the effect of this difference by  (i) calculating 
the reionization history of the universe, (ii)
fitting the observed UV luminosity functions of high redshift LBGs, and
(iii) calculating the feedback of star formation on the IGM.
The main results of our study are as follows:
\begin{itemize}

\item
In order to produce a given electron optical depth to the
reionization, the efficiency in the star formation and/or in
UV escape fraction has to be lower in the models using ST and other
mass functions obtained from simulations compared to that
with the PS mass function. This 
decrease has to be much more in the small mass molecular cooled halos
if they also contribute to the reionization process.

\item
All new sets of data points, which extend the UV luminosity function
of LBGs to the faint end and to higher $z$, can be naturally explained 
in the framework of our earlier models.

\item
The luminosity function determined using
the PS and ST mass functions match reasonably at $z=3$,
but they differ more and more strongly as one
goes to higher redshifts, and especially at the
bright end (right panel of Figure~\ref{fig_UV_lf}).
This is because as one moves to higher and higher redshifts
a given luminosity range is contributed by rarer and rarer 
objects and the PS mass function predicts a lower net formation rate
compared to the ST mass function or other mass functions
 derived from simulations.
The UV luminosity functions determined from all mass
functions, other than the PS mass function,
agree reasonably with each other.

\item
We find that the luminosity functions
calculated using the derivative of the PS mass function
and that obtained from the Sasaki formalism, differ negligibly
for $\kappa=1$ and by less than $30\%$ for other values 
of $0.1 < \kappa < 4$.

\item
{We show, by using $\chi^2$ minimization technique, that the
models with the ST halo mass function provide a better fit to
the observed UV luminosity functions in the redshift range
$3\le z \le 7.4$ compared to the models with the
PS halo mass function. However, the redshift evolution
of the best fit model parameter $f_*/\eta$ crucially
depends on the data set used, as well as
various uncertainties such as k-correction, dependence
of dust opacity on the wavelength etc.
}

\item
Models with different mass
functions, using the derivative of the mass functions to calculate
the net formation rate of halos, predict very similar
(with in 10\%) volume filling factor for metals in the IGM.
However, these models always predict a higher volume filling factor compared
to the Sasaki formalism of the PS mass function.
Therefore, 
even though the method used to calculate the formation rate
of collapsed dark matter halos has only a mild effect on 
the predictions of
high redshift UV luminosity functions, they do predict different
outflow feedback to the IGM. 
Different models give similar predictions for other physical
parameters associated with outflows and the conclusions of Paper II
appear to be largely insensitive to the adopted form of the halo mass
function.

\end{itemize}

The next important step in the development
of our models could be to implement it in the framework of
a numerical simulation or by generating merger
trees. This will help in unambiguously determining the halo
formation rate and its survival probability. 
However, as mentioned earlier,
one has to resolve the formation of halos right from 
$10^7M_\odot$ to $10^{12}M_\odot$ in order to model both
IGM metal enrichment and the galactic luminosity
function. To achieve such a large dynamic range
would be a major challenge.

\section*{acknowledgements}
We thank Iwata Ikuru
for kindly providing the data on
luminosity function at $z<6$.
We thank an anonymous referee for useful
suggestions and prompting us to do a $\chi^2$ analysis
of the UV luminosity functions.
SS thanks CSIR, India for the grant award
No. 9/545(23)/2003-EMR-I.

\end{document}